\documentclass{icrc2009}

\usepackage{graphicx}   
\usepackage[font=footnotesize]{subfig} 
\usepackage{fixltx2e}
\usepackage{url}

\newcommand{\shorttitle}[1]%
{\markboth{Proceedings of the 31\MakeLowercase{$^{st}$} ICRC, {\L}\'{o}d\'{z} 2009}{#1} }

\begin{document}
\title{Time Calibration of the Radio Air Shower Array LOPES}

\author{
\IEEEauthorblockN{
F.~Schr\"oder\IEEEauthorrefmark{1}, 
W.D.~Apel\IEEEauthorrefmark{1},
J.C.~Arteaga\IEEEauthorrefmark{2}$^,$\IEEEauthorrefmark{14},
T.~Asch\IEEEauthorrefmark{3},
F.~Badea\IEEEauthorrefmark{1},
L.~B\"ahren\IEEEauthorrefmark{4},
K.~Bekk\IEEEauthorrefmark{1}, 
}  \\ \vspace{-2.7ex}
\IEEEauthorblockN{
M.~Bertaina\IEEEauthorrefmark{5},
P.L.~Biermann\IEEEauthorrefmark{6},
J.~Bl\"umer\IEEEauthorrefmark{1}$^,$\IEEEauthorrefmark{2},
H.~Bozdog\IEEEauthorrefmark{1}
I.M.~Brancus\IEEEauthorrefmark{7},
M.~Br\"uggemann\IEEEauthorrefmark{8},
}  \\ \vspace{-2.7ex}
\IEEEauthorblockN{
P.~Buchholz\IEEEauthorrefmark{8},
S.~Buitink\IEEEauthorrefmark{4},
E.~Cantoni\IEEEauthorrefmark{5}$^,$\IEEEauthorrefmark{9},
A.~Chiavassa\IEEEauthorrefmark{5},
F.~Cossavella\IEEEauthorrefmark{2}, 
K.~Daumiller\IEEEauthorrefmark{1}, 
} \\ \vspace{-2.7ex}
\IEEEauthorblockN{
V.~de Souza\IEEEauthorrefmark{2}$^,$\IEEEauthorrefmark{15}, 
F.~Di~Pierro\IEEEauthorrefmark{5},
P.~Doll\IEEEauthorrefmark{1}, 
R.~Engel\IEEEauthorrefmark{1},
H.~Falcke\IEEEauthorrefmark{4}$^,$\IEEEauthorrefmark{10}, 
M.~Finger\IEEEauthorrefmark{1}, 
D.~Fuhrmann\IEEEauthorrefmark{11},
} \\ \vspace{-2.7ex}
\IEEEauthorblockN{
H.~Gemmeke\IEEEauthorrefmark{3},
P.L.~Ghia\IEEEauthorrefmark{9},
R.~Glasstetter\IEEEauthorrefmark{11}, 
C.~Grupen\IEEEauthorrefmark{8},
A.~Haungs\IEEEauthorrefmark{1}, 
D.~Heck\IEEEauthorrefmark{1}, 
} \\ \vspace{-2.7ex}
\IEEEauthorblockN{
J.R.~H\"orandel\IEEEauthorrefmark{4}, 
A.~Horneffer\IEEEauthorrefmark{4}, 
T.~Huege\IEEEauthorrefmark{1}, 
P.G.~Isar\IEEEauthorrefmark{1}, 
K.-H.~Kampert\IEEEauthorrefmark{11},
D.~Kang\IEEEauthorrefmark{2}, 
} \\ \vspace{-2.7ex}
\IEEEauthorblockN{
D.~Kickelbick\IEEEauthorrefmark{8},
O.~Kr\"omer\IEEEauthorrefmark{3},
J.~Kuijpers\IEEEauthorrefmark{4},
S.~Lafebre\IEEEauthorrefmark{4},
P.~{\L}uczak\IEEEauthorrefmark{12}, 
M.~Ludwig\IEEEauthorrefmark{2}, 
} \\ \vspace{-2.7ex}
\IEEEauthorblockN{
H.J.~Mathes\IEEEauthorrefmark{1}, 
H.J.~Mayer\IEEEauthorrefmark{1}, 
M.~Melissas\IEEEauthorrefmark{2}, 
B.~Mitrica\IEEEauthorrefmark{7},
C.~Morello\IEEEauthorrefmark{9},
G.~Navarra\IEEEauthorrefmark{5},
} \\ \vspace{-2.7ex}
\IEEEauthorblockN{
S.~Nehls\IEEEauthorrefmark{1},
A.~Nigl\IEEEauthorrefmark{4},
J.~Oehlschl\"ager\IEEEauthorrefmark{1}, 
S.~Over\IEEEauthorrefmark{8},
N.~Palmieri\IEEEauthorrefmark{2},
M.~Petcu\IEEEauthorrefmark{7}, 
T.~Pierog\IEEEauthorrefmark{1}, 
} \\ \vspace{-2.7ex}
\IEEEauthorblockN{
J.~Rautenberg\IEEEauthorrefmark{11}, 
H.~Rebel\IEEEauthorrefmark{1}, 
M.~Roth\IEEEauthorrefmark{1}, 
A.~Saftoiu\IEEEauthorrefmark{7}, 
H.~Schieler\IEEEauthorrefmark{1}, 
A.~Schmidt\IEEEauthorrefmark{3}, 
} \\ \vspace{-2.7ex}
\IEEEauthorblockN{
O.~Sima\IEEEauthorrefmark{13}, 
K.~Singh\IEEEauthorrefmark{4}$^,$\IEEEauthorrefmark{16}, 
G.~Toma\IEEEauthorrefmark{7}, 
G.C.~Trinchero\IEEEauthorrefmark{9},
H.~Ulrich\IEEEauthorrefmark{1},
A.~Weindl\IEEEauthorrefmark{1}, 
} \\ \vspace{-2.7ex}
\IEEEauthorblockN{
J.~Wochele\IEEEauthorrefmark{1}, 
M.~Wommer\IEEEauthorrefmark{1}, 
J.~Zabierowski\IEEEauthorrefmark{12},
J.A.~Zensus\IEEEauthorrefmark{6}
} 
\IEEEauthorblockA{\IEEEauthorrefmark{1}Institut f\"ur Kernphysik, Forschungszentrum Karlsruhe, Germany}
\IEEEauthorblockA{\IEEEauthorrefmark{2}Institut f\"ur Experimentelle Kernphysik, Universit\"at Karlsruhe, Germany}
\IEEEauthorblockA{\IEEEauthorrefmark{3}IPE, Forschungszentrum Karlsruhe, Germany}
\IEEEauthorblockA{\IEEEauthorrefmark{4}Department of Astrophysics, Radboud University Nijmegen, The Netherlands}
\IEEEauthorblockA{\IEEEauthorrefmark{5}Dipartimento di Fisica Generale dell' Universita Torino, Italy}
\IEEEauthorblockA{\IEEEauthorrefmark{6}Max-Planck-Institut f\"ur Radioastronomie Bonn, Germany}
\IEEEauthorblockA{\IEEEauthorrefmark{7}National Institute of Physics and Nuclear Engineering, Bucharest, Romania}
\IEEEauthorblockA{\IEEEauthorrefmark{8}Fachbereich Physik, Universit\"at Siegen, Germany}
\IEEEauthorblockA{\IEEEauthorrefmark{9}Istituto di Fisica dello Spazio Interplanetario, INAF Torino, Italy}
\IEEEauthorblockA{\IEEEauthorrefmark{10}ASTRON, Dwingeloo, The Netherlands}
\IEEEauthorblockA{\IEEEauthorrefmark{11}Fachbereich Physik, Universit\"at Wuppertal, Germany}
\IEEEauthorblockA{\IEEEauthorrefmark{12}Soltan Institute for Nuclear Studies, Lodz, Poland}
\IEEEauthorblockA{\IEEEauthorrefmark{13}Department of Physics, University of Bucharest, Bucharest, Romania}
\IEEEauthorblockA{\small \IEEEauthorrefmark{14}now at: Universidad Michoacana, Morelia, Mexico}
\IEEEauthorblockA{\small \IEEEauthorrefmark{15}now at: Universidade de S$\tilde{a}$o Paulo, Instituto de F\^{\i}sica de S$\tilde{a}$o Carlos, Brasil}
\IEEEauthorblockA{\small \IEEEauthorrefmark{16}now at: KVI, University of Groningen, The Netherlands}
}

\shorttitle{F.~Schr\"oder for LOPES - Time Calibration of LOPES}
\maketitle

\begin{abstract}
LOPES is a digitally read out antenna array consisting of 30 calibrated dipole antennas. It is located at the site of the KASCADE-Grande experiment at Forschungszentrum Karlsruhe and measures the radio emission of cosmic ray air showers in the frequency band from 40 to 80 MHz. LOPES is triggered by KASCADE and uses the KASCADE reconstruction of the shower axis as an input for the analysis of the radio pulses. Thereby LOPES works as an interferometer when the signal of all antennas is digitally merged to form a beam into the shower direction. To be sensitive to the coherence of the radio signal, a precise time calibration with an accuracy in the order of 1 ns is required. \\
Thus, it is necessary to know the delay of each antenna which is time and frequency dependent. Several calibration measurements are performed to correct for this delay in the analysis: The group delay of every antenna is measured regularly (roughly once per year) by recording a test pulse which is emitted at a known time. Furthermore, the delay is monitored continuously by the so called phase calibration method: A beacon (a dipole antenna) emits continuously two sine waves at 63.5 MHz and 68.1 MHz. By that a variation of the delay can be detected in a subsequent analysis of the radio events as a change of the phase at these frequencies. Finally, the dispersion of the analog electronics has been measured to account for the frequency dependence of the delay.
\end{abstract}

 
\section{Introduction}
 \begin{figure*}[th]
  \centering{\subfloat{\includegraphics[width=1.4in]{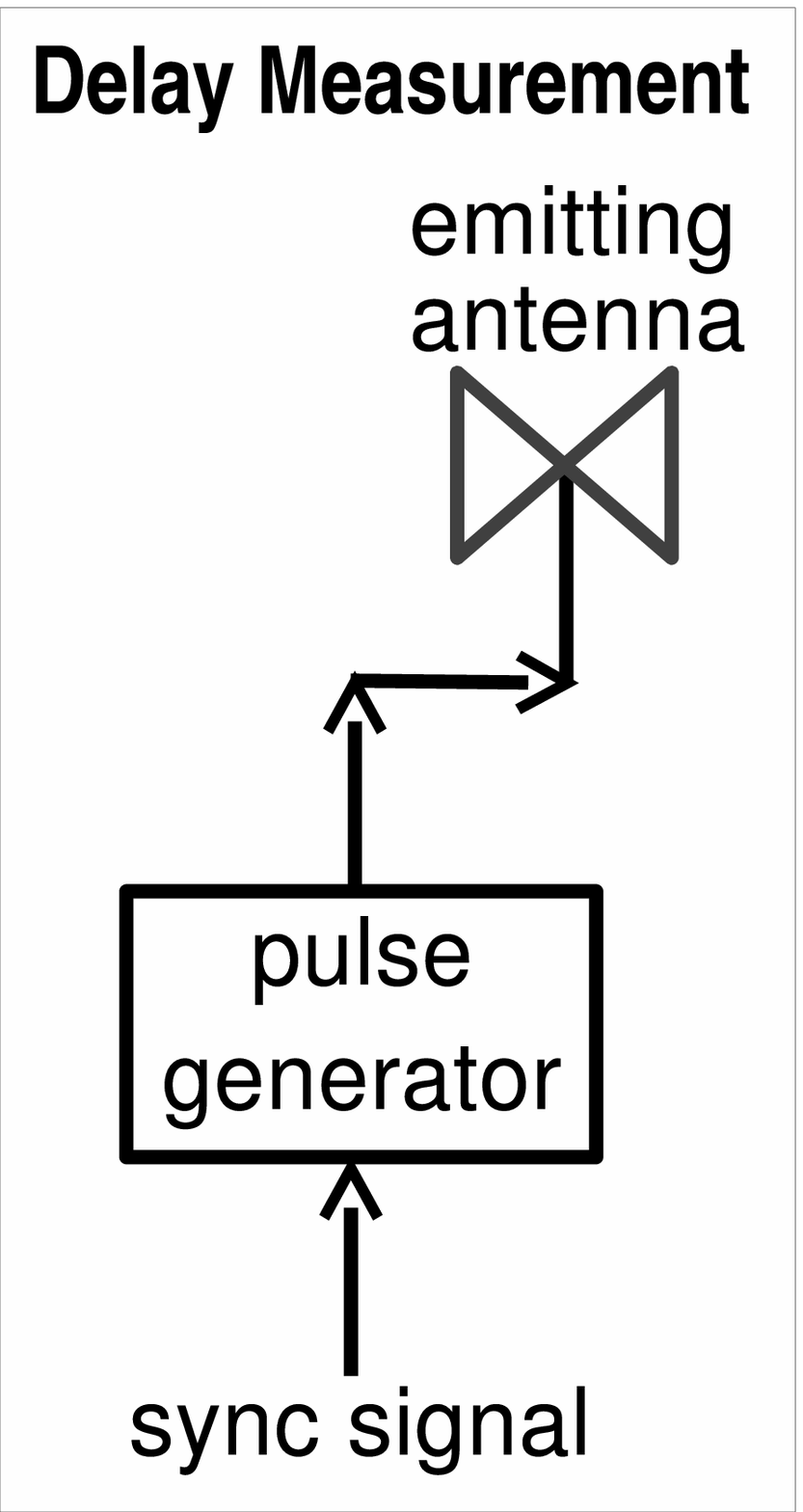}}
             \subfloat{\includegraphics[width=5in]{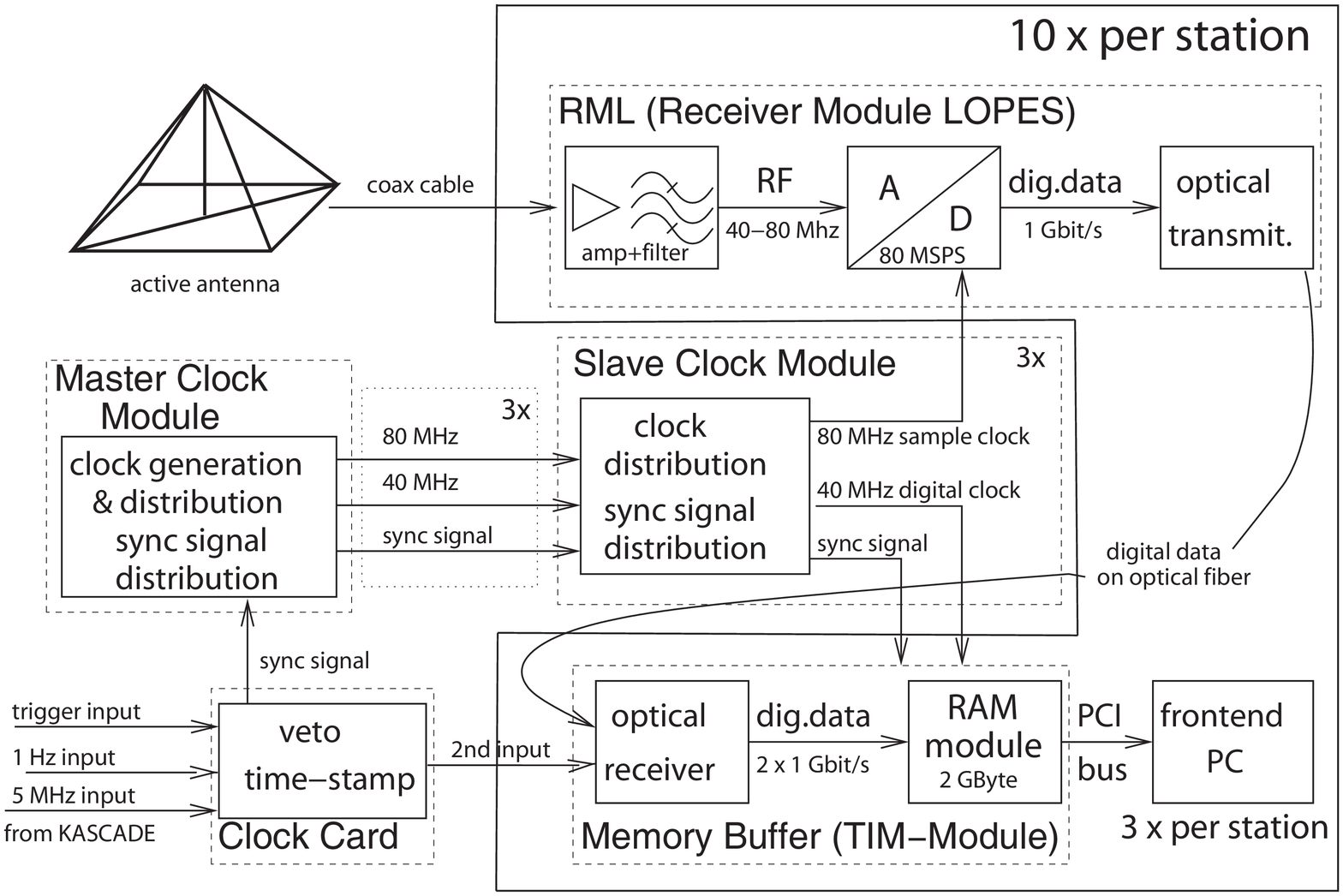}}
            }
  \caption{Right: LOPES hardware and timing setup. Left: extension used for delay measurements; a pulse generator which is connected to a emitting calibration antenna is triggered by the same sync signal as the LOPES DAQ and emits a short calibration pulse. Thus this calibration pulse is recorded with a certain delay within a normal event. Due to the setup, the delay of the pulse contains a certain offset, which is the same for every measurement and therefore can be ignored when determining relative delays between antennas.}
  \label{hardware_setup_fig}
 \end{figure*}
\noindent
The main part of LOPES (LOfar ProtoypE Station) consists of 30 digitally read out, absolutely calibrated, inverted V-shape dipole antennas \cite{Falcke05, Nehls08, Asch07}. The antennas are co-located with the KASCADE array with a baseline of about $200\,$m and are triggered about twice per minute by the KASCADE-Grande experiment \cite{Antoni03,Navarra04}. Only a few events per day contain a cosmic ray air shower radio pulse which is clearly distinguishable from the noise, as the noise floor is quite high inside the KASCADE array. Due to a precise time calibration of LOPES the digitally measured radio data can be used to form a beam into the shower arrival direction. Furthermore the cross-correlation of the antennas can be calculated to be sensitive to the coherence of the radio signal \cite{Horneff07}. This way LOPES is a phased array which can be used as a digital interferometer.

\noindent
Whenever LOPES is triggered a trace of $2^{16}\,$ samples with a sampling rate of $80\,$MHz is read out at each antenna with the trigger time roughly in the middle of the trace. LOPES is operating in the second Nyquist domain. Thus the full information of the radio signal between $40\,$MHz and $80\,$MHz is contained in the data and can be retrieved by up-sampling (i.e. the correct interpolation between the samples) with the zero-padding method \cite{Asch08}. With reasonable computing time, data can be up-sampled to a sample spacing below $0.1\,$ns, so that the sample spacing does not contribute significantly to uncertainties in the timing.

To obtain a stable timing of the antenna array the ADC clock is centrally generated and distributed via cables to all DAQ computers (fig.~\ref{hardware_setup_fig}). With the exception of jumps by full clock cycles (see section~\ref{sec_phasecalibration}), the jitter of the clock is negligible. Thus the time calibration of LOPES is basically reduced to measure the delay of each antenna and the subsequent electronics, i.e. the time between the arrival of a radio pulse at the antenna and its measurement with the DAQ. This delay is different for each antenna (mainly due to different cable lengths). As for interferometry only the differences of the arrival time of the radio signal between the antennas matters, the absolute delay is of minor importance. In this paper the term delay is therefore meant in a relative sense.

The following section explains how the delay is measured at LOPES. To achieve the necessary precision we also take into account second order effects, like the frequency dependence of the delay (dispersion) and variations of the delay with time.

\begin{figure*}[!t]
  \centerline{\subfloat[without dispersion correction]{\includegraphics[width=2.5in,angle=-90]{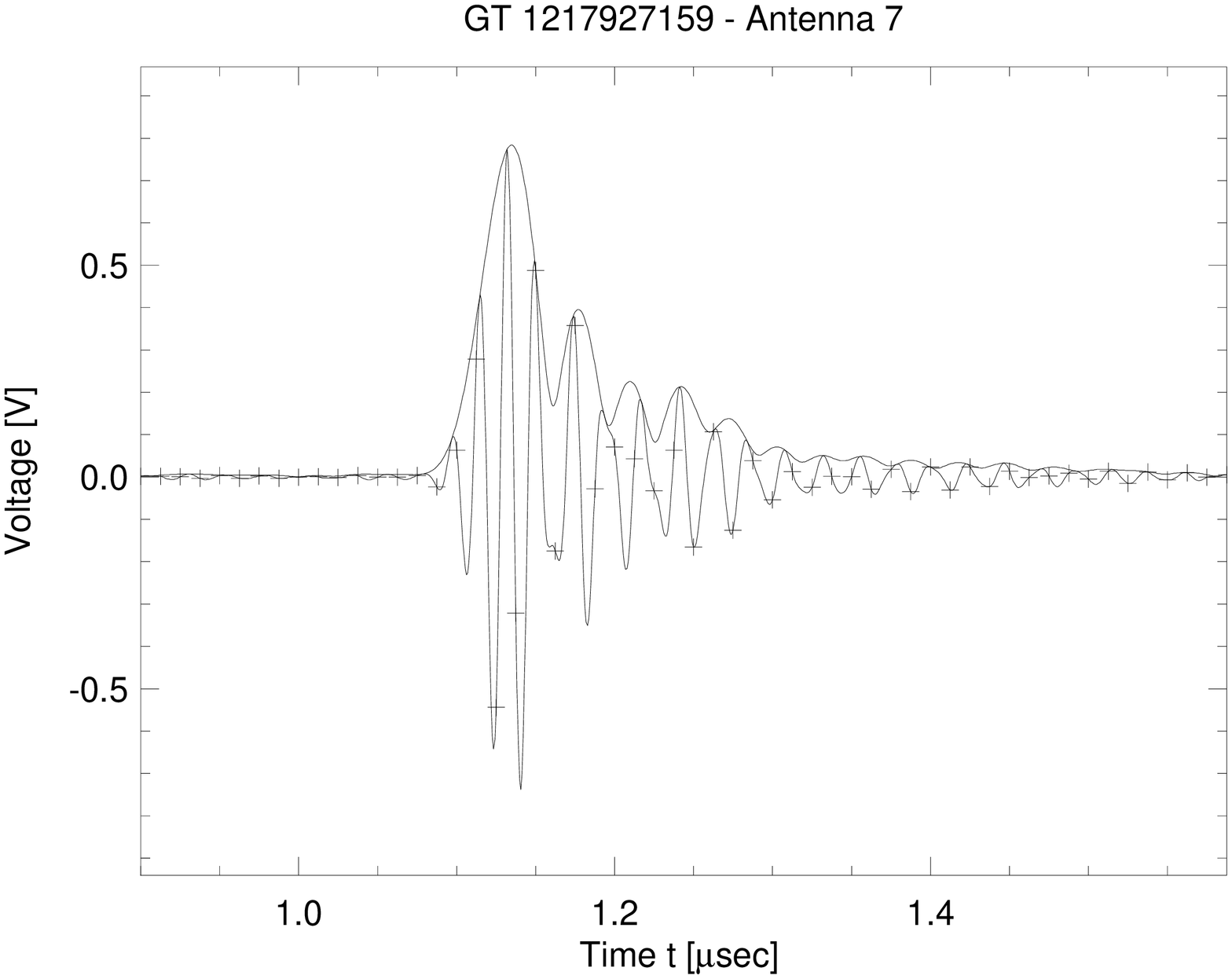} \label{sub_fig1}}
               \hfil
               \subfloat[after dispersion correction]{\includegraphics[width=2.5in,angle=-90]{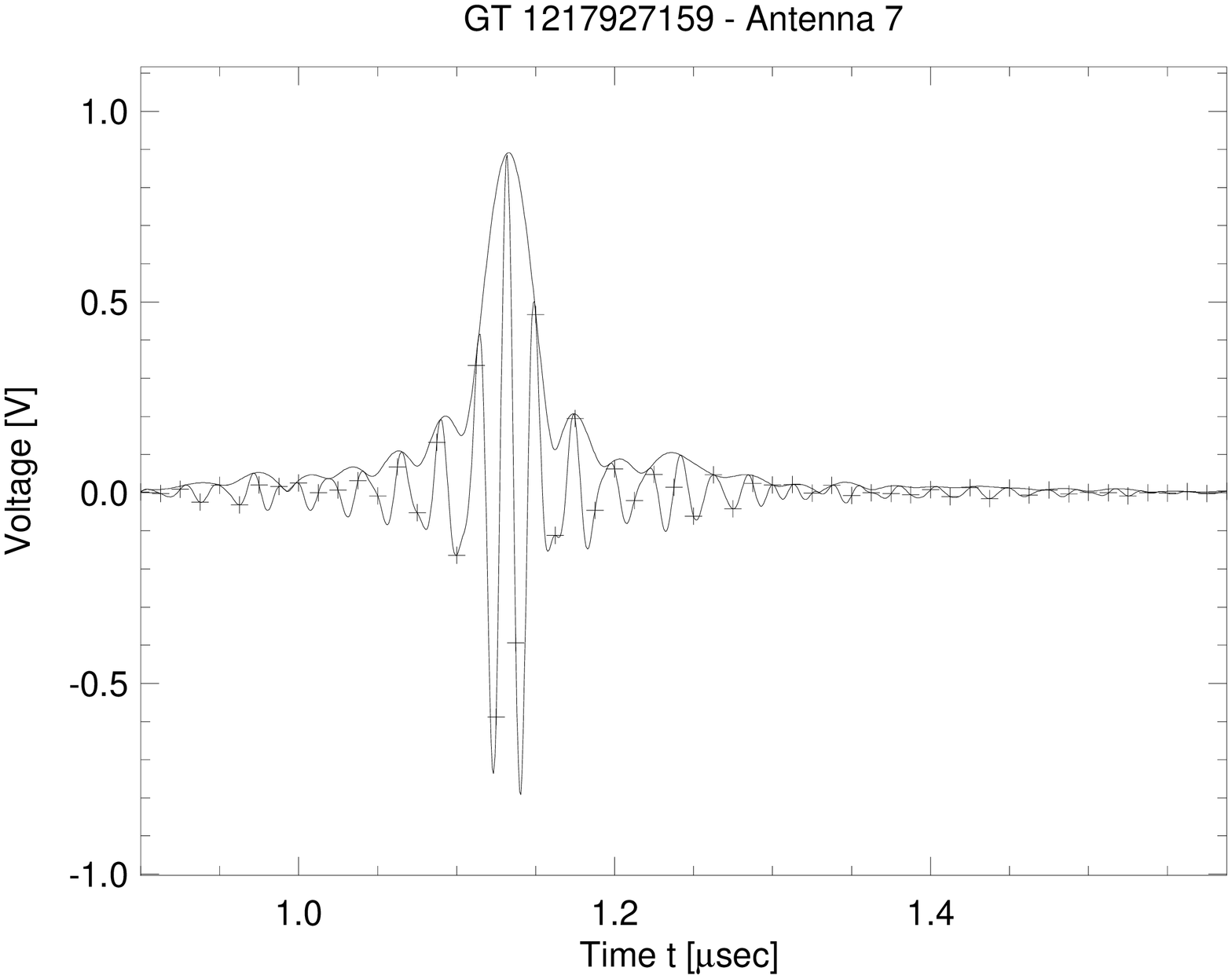} \label{sub_fig2}}
             }
  \caption{A short test pulse (FWHM $\approx 10\,$ ns) emitted by a function generator and recorded by the LOPES DAQ. The crosses indicate the measurement points sampled by the ADC. The lines show the up-sampled signal and the Hilbert envelope of the pulse. In the raw data (a) it can be seen that a part of the pulse is delayed by more than $100\,$ns due to the dispersion of the filter. After the correction of the dispersion (b), the pulse is more symmetrical, the position of the pulse envelope has moved by about $2\,$ns, its height (field strength) has increased by about $10\,$\% and its FWHM (field strength) decreased by about $10\,$\%. The effects on height and width are reduced to a few percent, when using the sub-band from $43\,$MHz to $74\,$MHz.}
   \label{fig_dispersion}
 \end{figure*}
\section{Delay measurements}
\noindent
LOPES as a digital interferometer requires the timing precision of the radio pulse in each antenna to be much smaller than the period of the filter ringing ($\approx 17\,$ns), when forming a cross-correlation beam into the arrival direction of the cosmic ray air shower. Thus the delay has to be known and measured with a precision of about $1\,$ns, to be sensitive to the coherence of the radio signal.

At the beginning of LOPES the delay has been measured with solar flare events, but now we have developed a new method which does not depend on any astronomical sources: Simultaneously with the LOPES data acquisition we trigger a pulse generator which is connected to a calibration antenna at a defined angle and distance to each LOPES antenna. The calibration antenna emits a pulse with a known delay after the trigger. Therefore after repeating the calibration for every antenna the pulse should appear in the data of each antenna at exactly the same time, if the delay of each LOPES antenna (and the connected analog electronics) would be the same.

However, e.g. due to different cable lengths, this is not the case, and indeed the calibration pulses are detected at different times for each antenna. The relative delay between different LOPES antennas can then be obtained by measuring the time differences of the detection times of the calibration pulses.

The pulse detection time can be determined by different methods: On the one hand a Hilbert envelope is calculated and the pulse position is taken as either the position of the maximum or the crossing of half height. Thereby the position of the maximum shows slightly less jitter (RMS about $0.4\,$ns) when looking at several events recorded within a few minutes and the delays obtained from the half height crossing and the maximum are consistent within errors. On the other hand the position can be defined as the minimum or the maximum of the up-sampled trace and the jitter for both of them for subsequent events is less than the used sampling spacing of $0.1\,$ns (after up-sampling). The delays calculated by the maximum and the minimum agree well with each other. Thus in principle the delay can be determined with a precision of better than $0.5\,$ns.

Although the delays determined by the two methods (pulse position by the trace itself or its envelope) are inconsistent by a few nanosecond, this inconsistency can be explained at least partially with the dispersion of the analog electronics. This difference is reduced to an average of about $2\,$ns when correcting for this dispersion. The remaining inconsistency could not be explained so far, and is under investigation. Nevertheless, since the calculation of the cross-correlation beam is done with the (up-sampled) trace and not with the envelope, we currently use the delays deferred directly from the trace for the analysis of cosmic ray events.

\section{Correction for the dispersion}
\noindent
The frequency dependence of the group delay of a system is called dispersion. For LOPES$^{STAR}$ the dispersion has been measured for the antenna, the analog electronics (filter) and the connecting cable \cite{Kroemer08}. The dispersion of the cable can be considered negligible and the largest contribution to the overall dispersion comes from the filter.

Unfortunately, the dispersion of the LOPES V-shape antenna is unknown, because it is not easily measurable. But the dispersion of the filter has been measured with a network analyzer and can be corrected in the analysis. This is especially important as the first 10 LOPES antennas are connected to a slightly different filter than the later 20 antennas. The successfull correction of the dispersion in the analysis software has been proven by recording test pulses which are emitted by a function generator connected to the RML (fig.~\ref{hardware_setup_fig}).

As expected, the filter creates a response only to the leading and falling edge of pulses. Thus a delta pulse with FWHM $<<$ filter width $^{-1}$ ($\approx 25\,$ns) should be seen as a pulse with a width of $\approx 25\,$ns (FWHM of the power). This pulse is linearly distorted by the dispersion such that it is partially delayed by more than $100\,$ns (fig.~\ref{fig_dispersion}). When correcting for the dispersion this distortion can be significantly reduced and the FWHM of the pulse power is in the order of $30\,$ns. In addition the pulse is shifted by a few ns and the pulse width and height (field strength) change by roughly $10 \%$. Thus the correction of the dispersion is necessary for precise time and amplitude measurements of radio pulses, and future radio experiments should aim to correct the dispersion of every system in the signal chain.

\section{Phase calibration} \label{sec_phasecalibration}
 \begin{figure}[!t]
  \centering
  \includegraphics[width=2.2in,angle=-90]{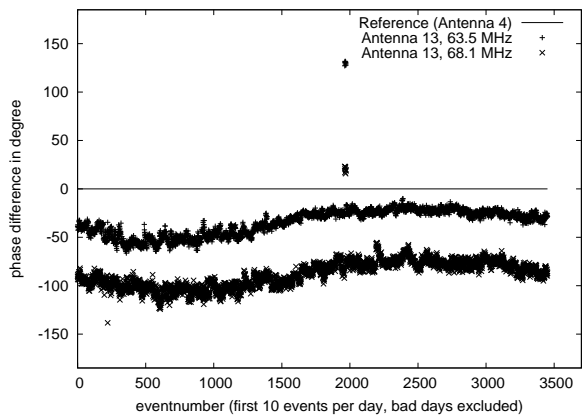}
  \caption{Example of the phase calibration: The phase differences at the beacon frequencies between one antenna and a chosen reference antenna are shown for the first 10 events of every day for one year of data taking (08 May 2008 - 07 May 2009, excluding a few days of down time and operational problems). }
  \label{phasecalibration_fig}
 \end{figure}
\noindent
The delay of each antenna and the read out electronics can be assumed to be roughly constant. However, there are variations on the scale of a few ns over time. In addition the LOPES clock distribution is not absolutely temperature stable, causing the effective delay of single antennas to increase or decrease from time to time by integer steps of $12.5\,$ns (one clock cycle). To correct for these steps, originally the phase of a TV transmitter inside of the LOPES band was monitored. After the shut down of this TV transmitter this method has been further developed by setting up a dedicated transmitter (beacon) inside of the Forschungszentrum Karlsruhe.

The beacon continously emits two sine waves of $63.5\,$MHz and $68.1\,$MHz at $-21\,$dBm each, which have a very narrow band width (FWHM $< 100\,$Hz). This signal is clearly seen above the noise level by all LOPES antennas and the phase can be measured at each of the two beacon frequencies at each antenna. For each of the two frequencies now the following is valid: As a LOPES event contains data from every antenna coincidently, a stable delay should lead to a constant difference between the phases at the same frequency measured by different antennas. Thus a variation in the relative delay between two antennas can be detected as a variation in the phase difference at the same two antennas at one of the beacon frequencies. Subsequently this variations can be taken into account in the analysis of radio events.

In a few events the measurement of the phase is disturbed by RFI noise. To avoid corrections for \rq unreal\lq~variations of the delay and to take into account ambiguities of the phase if the difference is larger than $180^{\circ}\,$,a consistency check is performed between the phase differences at both frequencies. For most of the events the results agree well, and this way the timing of the antennas can be monitored and significantly improved on a event-by-event basis.

For the given example (fig.~\ref{phasecalibration_fig}), the correlated drift which can be seen in both frequencies corresponds to a variation in the effective delay of about $1.5\,$ns over the year and can be clearly distinguished from the jitter of the phase differences. The jump in the middle is caused by a change of the effective delay by $25\,$ns during one day. The outlier at the beginning is one of the few noisy events, for which the phase calibration method would fail.

The accuracy of the phase calibration on short time scales is thereby determined by the short-term jitter (noise) of the phase measurement. This jitter is a function of the trace length and the amplitude of the beacon signal. Therefore, for a fixed trace length ($2^{16}$ samples at LOPES) an emission power of the beacon can be chosen such that the phase calibration allows a sufficiently accurate correction of the timing. In the case of LOPES the short term jitter of the phase differences is in the order of $0.3\,$ns. Furthermore, in figure \ref{phasecalibration_fig} can be seen that there is some additional error on longer time scales, as changes in the phase differences are not completely correlated between both frequencies. But still, the overall accuracy is better than the required timing precision of about $1\,$ns.

\section{Conclusions}
\noindent
It could be shown that the delay of each LOPES antenna and the corresponding electronic chain can be calibrated with a precision of better than $0.5\,$ns. Nevertheless, there is a not fully understood difference between the measurement of the time of a radio pulse by either looking to the trace or its envelope. In addition, it is possible to continously monitor the relative delay of each antenna with a precision of below $1\,$ns. Thus long and short term variations of the delay in the order of a few nanoseconds can be corrected in the subsequent data analysis on an event-by-event basis.

Furthermore, the dispersion of the analog electronics has been shown to slightly affect the time, the width and the height of measured radio pulses. Therefore this is taken into account in the LOPES standard analysis pipeline to produce more accurate results. Also future radio experiments should consider the dispersion of the whole detection system.

Finally, the different timing calibration methods of LOPES show that radio air shower experiments can achieve a relative timing accuracy of better than one nanosecond and that this accuracy can be monitored and maintained continuously over long periods.


\begin{thebibliography}{99}
   \bibitem{Falcke05}   H. Falcke et al., \emph{Nature 435 (2005) 313}
   \bibitem{Nehls08}    S. Nehls et al., \emph{Nucl.Instr.and.Meth.A 589 (2008) 350}
   \bibitem{Asch07}     T. Asch et al., \emph{Proceedings of the 30th International Cosmic Ray Conference, Merida, Mexico 5 (2008) 1081}
   \bibitem{Antoni03}   T. Antoni et al., \emph{Nucl.Instr.and.Meth.A 513 (2003) 429}
   \bibitem{Navarra04}  G. Navarra et al., \emph{Nucl.Instr.and.Meth.A 518 (2004) 207}
   \bibitem{Horneff07}  A. Horneffer et al., \emph{Proceedings of the 30th International Cosmic Ray Conference, Merida, Mexico 4 (2008) 83}
   \bibitem{Asch08}     T. Asch, \emph{FZKA report 7459, Forschungszentrum Karlsruhe (2009)}
   \bibitem{Kroemer08}  O. Kr\"omer, \emph{FZKA report 7396, Forschungszentrum Karlsruhe (2008)}
  \end{thebibliography}
\end{document}